\input{aipcheck}

\documentclass[
    ,final            
  ]
  {aipproc}

\layoutstyle{6x9}

\newcommand{\degr}{^{\rm o}}
\begin{document}

\title{Large-scale coherent orientations \\of quasar polarisation vectors: \\interpretation in terms of axion-like particles}

\classification{
		14.80.Mz,
		95.30.Gv,
		98.54.Aj,
		98.65.Dx
	       }
\keywords      {
		axion,
		polarisation,
		wave-packets,
		active galactic nuclei,
		large-scale structure of the Universe
		}

\author{A.~Payez}{
	address={University of Li\`{e}ge, All\'{e}e du 6 Ao\^{u}t 17, 4000 Li\`{e}ge, Belgium}
}
\author{D.~Hutsem\'ekers}{
	address={University of Li\`{e}ge, All\'{e}e du 6 Ao\^{u}t 17, 4000 Li\`{e}ge, Belgium}
}
\author{J.R.~Cudell}{
	address={University of Li\`{e}ge, All\'{e}e du 6 Ao\^{u}t 17, 4000 Li\`{e}ge, Belgium}
}

\begin{abstract}
The observation of redshift-dependent coherent orientations of quasar polarisation vectors over cosmological distances in some regions of the sky is reviewed. Based on a good-quality sample of 355 measured quasars, this observation seems to infer the existence of a new effect acting on light propagation on such huge distances.

A solution in terms of nearly massless axion-like particles has been proposed in the literature and its current status is discussed.
\end{abstract}

\maketitle


\section{The observations}

Observational evidence for large-scale alignments of quasar
polarisation vectors is discussed in a series of papers
\cite{hut98,hut01,hut05,jai04}.  The most recent sample contains 355
polarised quasars up to redshifts $z \sim 2.5$ and with optical
($\lambda \sim 550$ nm) polarisation angles ($\theta$)
determined with uncertainties $\sigma_{\theta} \leq 14\degr$.
Instrumental polarisation was measured to be very small and quasar
polarisation data obtained at different observatories with different
instruments do agree within the uncertainties. To minimize
contamination by interstellar polarisation in our Galaxy, only objects
with polarisation degrees $p \geq 0.6\%$ and located at high ($\geq
30\degr$) Galactic latitudes were considered. The sample contains
various types of quasars: radio-quiet, radio-loud, with or without
absorption lines. As far as possible, blazars were excluded due to
their variable polarisation and unsecure redshifts.

\begin{figure}[]
  \includegraphics[width=0.971\textwidth]{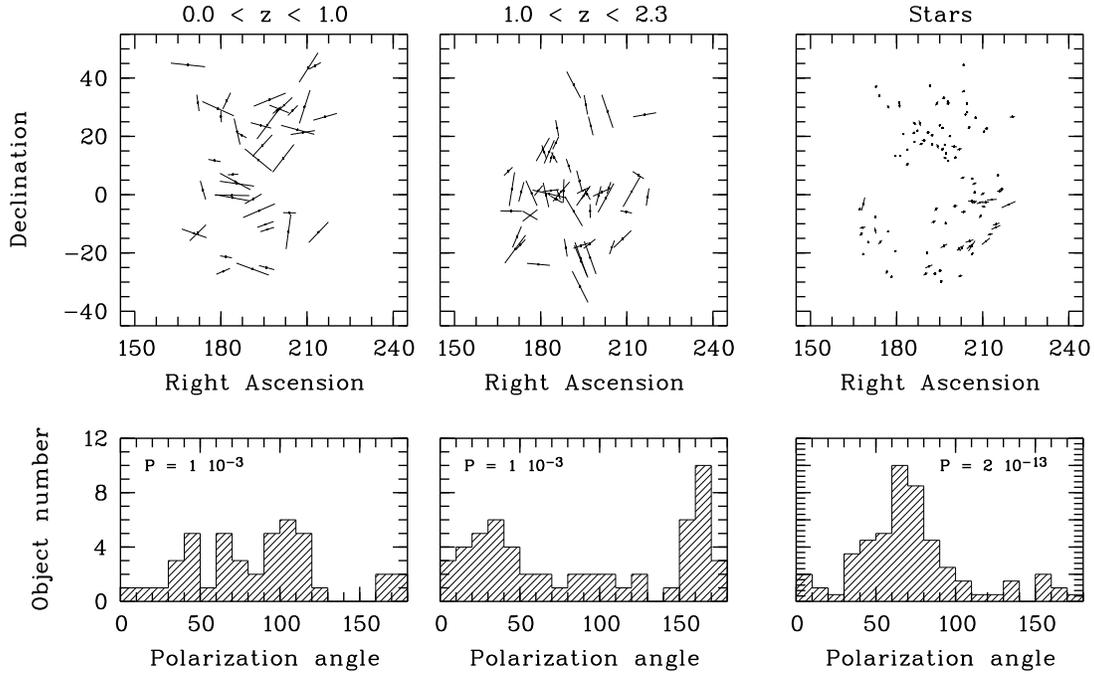} 
  \label{fig:fig1}
  \caption{The upper panels represent maps in equatorial coordinates
in which polarised objects are plotted: low-redshift ($z \leq 1$)
quasars (left), high-redshift ($z \geq 1$) quasars (middle), and the
Galactic stars from \cite{hei00} angularly closest to the quasars
(right).  The length of the polarisation vectors is proportional to
the polarisation degree $p$ below $2\%$, and constant above this
value to avoid confusion. The lower panels represent the polarisation
angle distributions ($180\degr$ is equivalent to $0\degr$). The
probability P that the polarisation angles are uniformly distributed
is given in each panel. }
\end{figure}

The alignment effect is illustrated in Fig.~\ref{fig:fig1} which shows
maps of polarisation vectors in a selected region of the sky.  Quasars
at low ($z < 1$) and high ($z > 1$) redshifts are separated.  We
immediately see that the polarisation vectors are not randomly
oriented, despite what one might expect.  Statistical tests designed for circular
data (angles) indicate that the departure from an isotropic
distribution of polarisation angles is significant (we used the
Hawley-Peebles test \cite{haw75}, with bins $\Delta\theta= 30\degr$ to
have a sufficient number of objects per bin). Moreover, the
polarisation angles of low- and high-redshift quasars appear to
cluster around significantly different mean directions ($< \theta > =
79\degr$ and $< \theta > = 8\degr$, respectively). The sources cover a
huge region of the sky, typically 1$\,$Gpc at $z \sim 1$.

\begin{figure}[]
  \includegraphics[trim = 0mm 0mm 0mm 0.5mm, clip, width=0.971\textwidth]{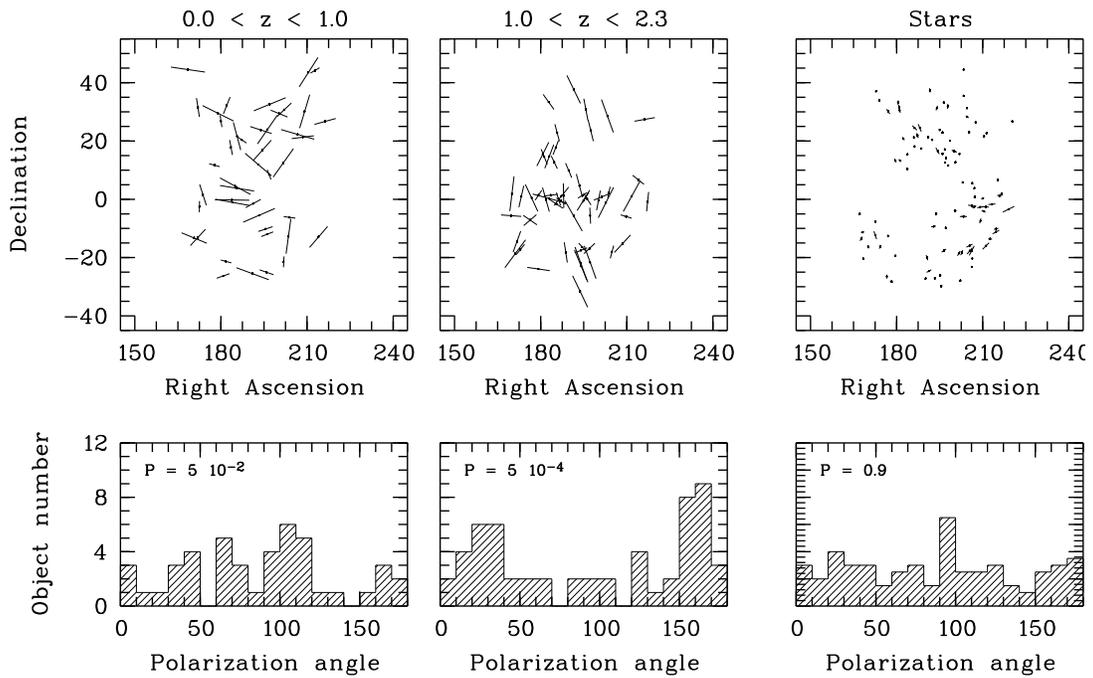}
  \label{fig:fig2}
  \caption{Same as Fig.~\ref{fig:fig1}, but the polarisations
are corrected for interstellar polarisation using a model built from
Galactic star polarisations (right panels of Fig.~\ref{fig:fig1}).}
\end{figure}

\begin{figure}[]
  \includegraphics[trim = 0mm 0mm 0mm 0.5mm, clip,width=0.971\textwidth]{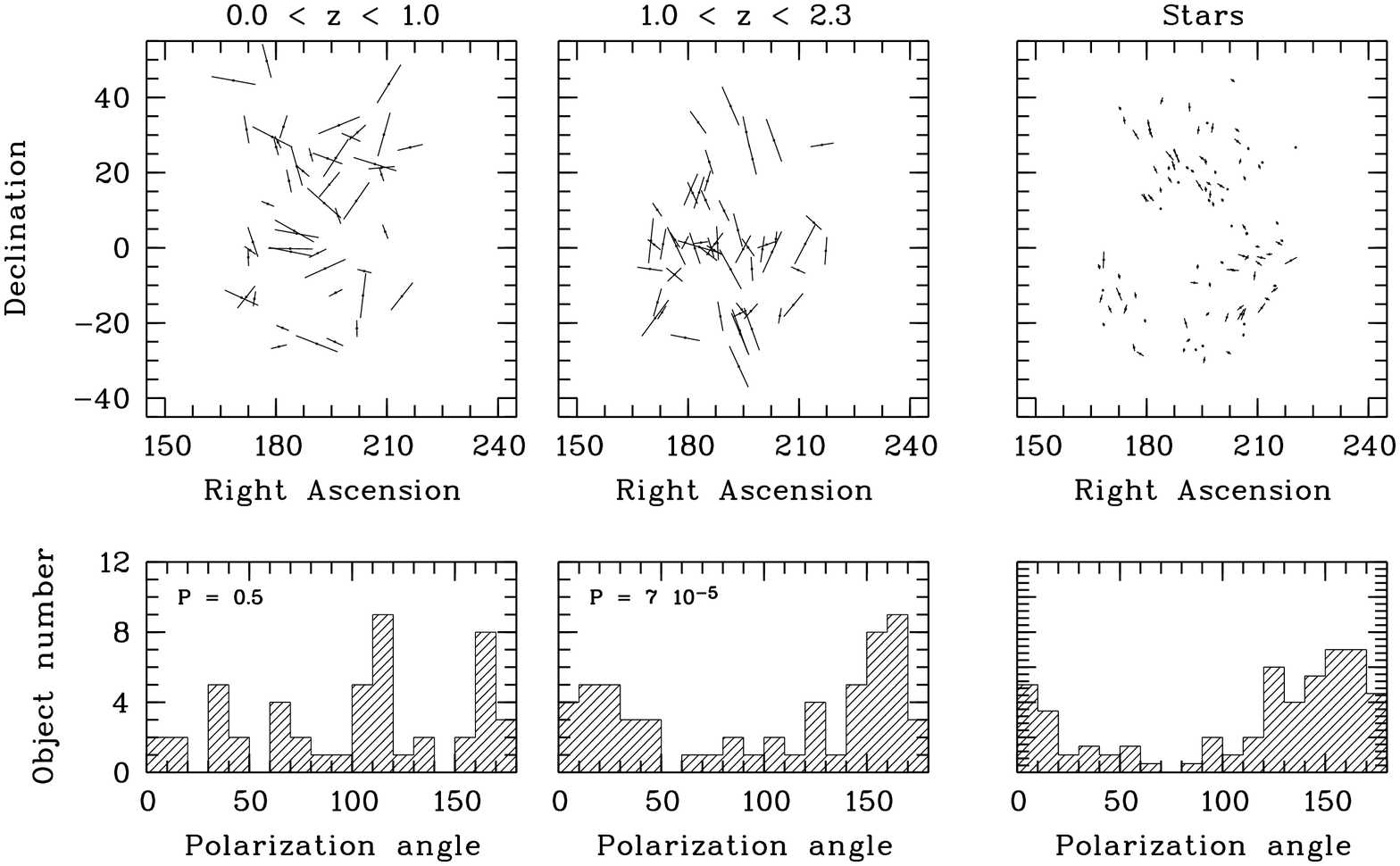}
  \label{fig:fig3}
  \caption{Same as Fig.~\ref{fig:fig2}, but the polarisations are
overcorrected up to the quasi-randomisation of the low-redshift quasar
polarisation angles.}
\end{figure}

At high galactic latitudes the interstellar polarisation is typically
$\leq 0.2-0.3\%$ (Fig.~\ref{fig:fig1}) but can occasionally be higher.
Although we adopt the cutoff $p \geq 0.6\%$ on the polarisation degree
to make sure that most of the measured polarisation is intrinsic to
the quasars, the polarisation of the objects in our sample is usually
lower than 2\% so that at least some data might be affected by
interstellar polarisation.  Several tests and simulations were
performed \cite{hut05,cab05,slu05} leading to the conclusion that it
is very unlikely that interstellar polarisation can be at the origin
of the observed alignments. This is also illustrated in
Figs.~\ref{fig:fig2} and~\ref{fig:fig3}. First, we model the
interstellar polarisation with a constant polarisation degree and a
gaussian distribution of polarisation angles which roughly reproduces
the Galactic star polarisation seen in Fig.~\ref{fig:fig1}. Then, we
subtract the modeled polarisation from both the quasar and the stellar
polarisations. As seen in Fig.~\ref{fig:fig2}, this correction has
little effect on the quasar polarisation vector alignments which
remain significant. In Fig.~\ref{fig:fig3}, we exaggerate this
correction up to a quasi-randomization of the polarisation angles of
low-redshift quasars. This results in an enhancement of the alignments
at high-redshift, demonstrating that it is not possible to
simultaneously wash out the effect at both low and high redshift.  In
other words, should interstellar polarisation (or any local
contamination) be responsible for the observed alignments, one would
expect the same effect at all redshifts and not a significant
difference between the mean polarisation angles at low and high
redshifts.

Global statistical tests based on the nearest-neighbor analysis were
applied to the full sample of 355 quasars.  The basic idea is to
compute for each quasar a statistics which measures, for instance, the
dispersion of the polarisation angles for a group of $n_{\rm v}$
neighbors.  Statistics can be calculated for the entire sample and the
significance of a departure from isotropy assessed with a Monte-Carlo
reshuffling of the polarisation angles over 3D positions. The
significance level is then computed as the probability that a random
configuration has a more extreme statistics than the observed one.  The
results indicate that the quasar polarisation vectors are not randomly
distributed over the sky, with a probability in excess of 99.9\%
\cite{hut05,jai04}. Coherent orientations are best detected in groups
of 30-40 objects which correspond to the regions of alignments seen in
Fig.~\ref{fig:fig1}.

To summarize, the observations show the existence of large-scale
alignments of quasar polarisation vectors over huge regions of the sky.
The alignments are significant and robust against contamination by
interstellar polarisation. They are observed at optical wavelength and
not at radio wavelengths \cite{jos07}, which might provide clues to
their origin.

\section{An axion--photon mixing interpretation}

Knowing the properties of axion--photon mixing~\cite{sikivie,raffeltstodolsky}, especially its effect on polarisation (namely \emph{dichroism} and also \emph{birefringence}, see e.g.~\cite{reviewHLPW}), it is tempting to consider this as an explanation for the discussed observations; the idea being that this mixing would take place on the way from quasars to Earth, inside extragalactic magnetic fields.
Indeed, let us recall that dichroism is known as an appealing way to create linear polarisation, as, in an external magnetic field $\vec{\mathcal{B}}$ (which then defines a particular direction), axion-like particles (ALPs) only couple to one polarisation of light ---parallel (resp. perpendicular) to $\vec{\mathcal{B}}$ in the case of pseudoscalar (resp. scalar) ALPs.

\begin{center}
		\begin{figure}
			\includegraphics[width=0.75\textwidth]{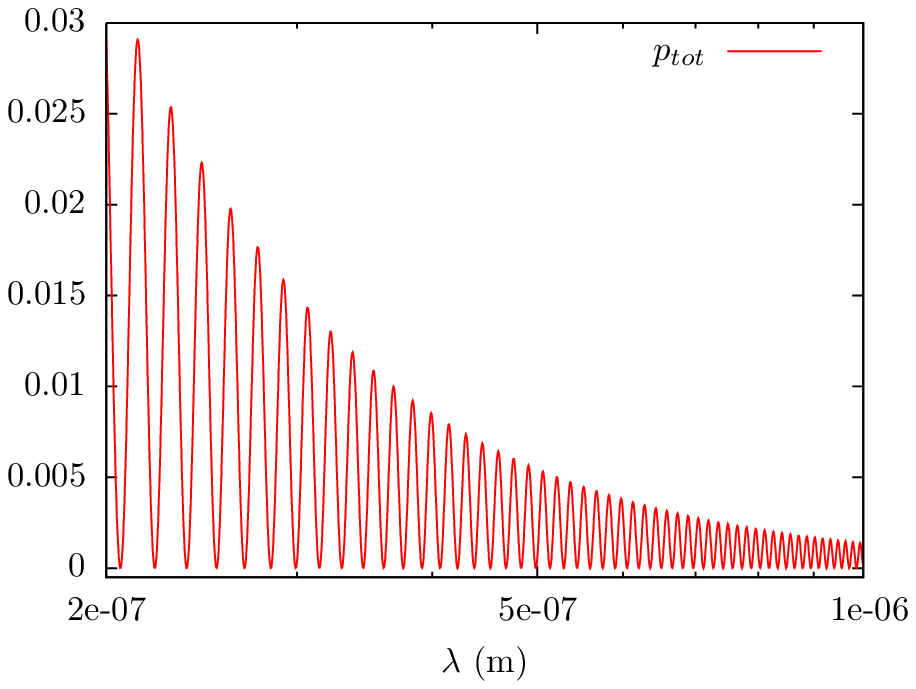}
			\caption{Degree of polarisation purely generated by axion--photon mixing, as a function of the wavelength, for initially unpolarised monochromatic light beams. Parameters: ALP mass, $m = 4.5~10^{-14}$~eV; plasma frequency, $\omega_p = 3.7~10^{-14}$~eV; extension of the magnetic field region, $L=10$~Mpc; coupling, $g|\vec{\mathcal{B}}|=2.17$~Mpc$^{-1}$ (e.g., $g=7~10^{-12}$~GeV$^{-1}$, $|\vec{\mathcal{B}}|=0.1~\mu$G). 
			}
			\label{fig:pol}
		\end{figure}
	\end{center}

Following this idea, it has been checked by several authors that it is possible to create the amount of linear polarisation needed to mimic an alignment effect, similar to the one observed, if one considers very light axion-like particles ($m\sim10^{-14}$~eV, $g\sim10^{-12}$~GeV$^{-1}$). This additional amount of linear polarisation has been estimated to be between 0.5\% and 2\% and is already achievable (Fig.~\ref{fig:pol}) in the simple case of a single magnetic field region (e.g., $\mathcal{B} = 0.1~\mu$G, $L=10$~Mpc~\cite{vallee}), see e.g.~\cite{reviewHLPW,das,planewaves}. Also, the energy-dependence of the mixing agrees with the fact that the alignment is not found at radio wavelength: indeed, the higher the energy of the photon, the more efficient the axion--photon mixing.

What remains to be done is thus to imagine a realistic way to create two different preferred directions, depending on the redshift. While two transversally huge magnetic field regions roughly located at redshift $\sim0$ and redshift $\sim1$ should, quite trivially, be able to produce a similar effect, the existence of such a ``cosmic sandwich'' seems unlikely, and a more refined description is needed. A step in that direction is found, for example, in the paper~\cite{mandal}, in which the possibility of creating an axion flux large enough for distant quasars is discussed\footnote{Even though the details on how to estimate the dimming of photons might differ.} and in which part of these ALPs convert back into photons in a closer magnetic field region and lead to a different preferred direction. 

However, new data (which are still under analysis~\cite{polcirc}) have been obtained and might be in serious conflict with the usual way this kind of ALP solution is presented, as they indicate that the amount of circular polarisation of light coming from quasars is significantly smaller than the linear one. Referring to what is predicted by birefringence, this is actually not what one would naively expect from axion--photon mixing, when describing light with monochromatic plane-waves. Indeed, the general result in that case is that the linear polarisation and the circular polarisation are expected to be of the same order
. Should this be an unavoidable prediction of the mixing, the explanation of the large-scale coherent orientations of quasar polarisation vectors in terms of ALPs would be discarded. Nevertheless, this is not the case, as, in a more realistic formalism, using wave-packets, it can be shown that the circular polarisation produced is significantly reduced \cite{paper,proceedpolcirc}. These new observations thus only constrain the axion--photon mixing, which can still be considered to interpret the reported alignments.

Finally, it worth keeping in mind that similar nearly massless axion-like particles have also been considered to address different astrophysical observations. In \cite{roncadelli}, the authors try to explain the high degree of transparency of the Universe and also the spectrum of a particular blazar (3C279), using such ALPs. Interestingly enough, this blazar is also part of the very sample of 355 polarised quasars presented here, for which we have considered axion-photon mixing to discuss the polarisation. These particles have also been discussed by other people in the context of cosmic rays~\cite{fairbairn} or of AGN luminosity relations~\cite{burrage}, for example.

\section{Conclusion}

We have reviewed the observations of large-scale alignments of quasar polarisation vectors
.
The contribution of interstellar polarisation has been shown to be negligible and the existence of preferred directions, depending on the redshift, has been proved to be robust.

The current status of an explanation in terms of the mixing of the incoming photons with nearly massless axion-like particles in external magnetic fields has then been discussed; the main missing ingredient being the size and the configuration of the magnetic fields on the line of sight needed to reproduce the whole set of observations.


\begin{theacknowledgments}

A. P. would like to thank Fredrik Sandin, Davide Mancusi and Javier Redondo for useful discussions on physical and numerical matters. We also acknowledge interesting discussions with Subhayan Mandal.

\end{theacknowledgments}



\bibliographystyle{aipproc}   




\end{document}